# Removing Gamification: A Research Agenda

**KATIE SEABORN**

*Department of Industrial Engineering and Economics*
*Tokyo Institute of Technology*

# Removing Gamification: A Research Agenda

**Katie Seaborn**
Tokyo Institute of Technology
Tokyo, Japan
seaborn.k.aa@m.titech.ac.jp

## ABSTRACT

The effect of removing gamification elements from interactive systems has been a long-standing question in gamification research. Early work and foundational theories raised concerns about the endurance of positive effects and the emergence of negative ones. Yet, nearly a decade later, no work to date has sought consensus on these matters. Here, I offer a rapid review on the state of the art and what is known about the impact of removing gamification. A small corpus of 8 papers published between 2012 and 2020 were found. Findings suggest a mix of positive and negative effects related to removing gamification. Significantly, insufficient reporting, methodological weaknesses, limited measures, and superficial interpretations of "negative" results prevent firm conclusions. I offer a research agenda towards better understanding the nature of gamification removal. I end with a call for empirical and theoretical work on illuminating the effects that may linger after systems are un-gamified.

## CCS CONCEPTS

• Human-centered computing

## KEYWORDS

Gamification, Longitudinal effects, Gamification removal, Un-gamify

## 1 INTRODUCTION

Gamification is the use of game elements in non-game systems to provide a game-like experience and influence user motivation and engagement [8,19,40]. As an interdisciplinary topic, gamification has been explored across a variety of domains and areas of practice, from education to health to marketing and more [17,22,32,40]. The field of gamification research is maturing. Since the first reviews of the literature conducted about 5-6 years ago, there has been a dramatic increase in studies and surveys on gamification. The ACM Digital Library, for instance, showed a 2¼-fold increase in gamification-related work between 2016 and the present[1]. Much of this body of work has addressed the gaps, issues, and opportunities originally identified with increasing complexity, specificity, and rigor.

Yet, one critical question raised early on through foundational studies [42], theoretical predictions [4,5], and previous surveys [40] has remained open: the effect of *removing* gamification, or "un-gamifying" a system. In 2011, Cunningham and Zichermann [4] noted that the most common approaches to gamifying systems targeted extrinsic motivation, or motivation that is driven by external rewards rather than intrinsic satisfaction [5]. As suggested by the formative work of Deci, Ryan, and colleagues [5,37,38], extrinsic motivation may not lead to lasting behavioral change after the rewarding mechanism is removed. Thus, by using game elements that tend to act as extrinsic incentives, like points [26,27] and leaderboards [21,41], users tend to be extrinsically engaged rather than finding the system intrinsically enjoyable. As such, the system may need to stay gamified forever—otherwise, the positive effects may be lost, or even turn negative. In their seminal 2012 paper, Thom et al. [42] addressed this directly by designing a study wherein gamification was removed after a 10-months period. They found that participation drastically decreased. This negative result was featured by Seaborn and Fels [40] in their 2015 survey paper, where they called for studies on removing gamification in their research agenda. In 2017, as the field started to mature, the dire need for deeper studies on such behavioral patterns and their connection to theory was echoed by Nacke and Deterding [31] in their research agenda. Most recently, Rapp et al. [35] signaled a need for ethical reflection on integrating gamification into (and presumably out of) people's lives.

At nearly a decade after the initial work and five years after the first call, it is time to trace out what we know about gamification removal. The stakes are high given the sheer amount of gamification work conducted in CHI and other spaces. If there are negative effects upon removing gamification, they could be widescale, and if not reported, they could be invisible. Thus, the objective of this rapid review was to understand the state of the art on gamification removal. I asked: *What is the nature of the research and findings on removing gamification?* The contributions are twofold: (1) to identify and describe how the question of gamification removal has been addressed in HCI; and (2) to provide a consensus on the effects of gamification removal. Taken together, this work extends theoretical and empirical knowledge in the field of gamification on a key topic with widespread and continuing implications.

---

[1] Calculated from a search using the keyword "gamification" conducted on December 15, 2020, which returned 892 results between 2011 and 2015, inclusive, and 2005 results between 2016 and 2020, inclusive.

## 2 METHODS

A rapid review [9,11,13,24] was conducted. Rapid reviews are appropriate for focused, well-defined, and time-sensitive research questions that produce descriptive statistics, narrative summaries of qualitative data, and possibly some quantitative findings. The research question posed for this work is well-suited for a rapid review, being narrowly defined, timely, and geared towards producing a descriptive and narrative synthesis of the surveyed literature. A version of the PRISMA checklist [25] adapted[2] for HCI purposes was used to guide the rapid review process and structure the reporting of results. See Figure 1 for the PRISMA flow chart. I conducted all database searches, data extraction, and data analyses alone, which is common and adequate for rapid reviews [24]. For greater rigor and accountability, this protocol (OSF # 75cqh) was registered before data extraction on December 15, 2020[3].

### 2.1 Eligibility Criteria

Full studies involving human subjects interacting with a gamified computer system were included; pilot studies, proposals, technical reports, surveys, and grey literature were excluded. Only uses of gamification in sense defined above—game elements rather than full games—were included; serious games, gameful design, and so on were excluded. Only studies that included the removal of gamification features and evaluated the effects of doing so were included. Inaccessible studies, such as offline work, and studies not in English were excluded.

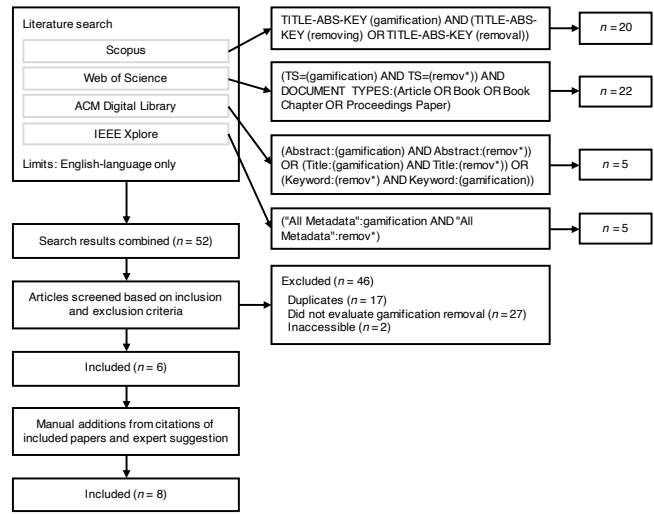

**Figure 1: PRISMA flow diagram of the rapid review procedure.**

### 2.2 Search Queries and Study Selection

Queries were run in general databases (Scopus and Web of Science) and disciplinary databases (ACM Digital Library and IEEE Xplore) on December 15, 2020. All queries used the terms "gamification" and "remov*" or, when it was not possible to use the * qualifier, a combination of "removal" and "removing." Duplicates were removed. The abstract and, if necessary, full text was then reviewed against the eligibility criteria. The related work sections of the selected papers and the citing papers were manually reviewed for additional relevant papers. An outside expert suggested an additional paper. In the end, from an initial 52 papers, 7 were selected for inclusion.

### 2.3 Data Analysis

I extracted all data and metadata, including: the study type, research approach (quantitative, qualitative, mixed), lab or field, timeframe of the study, area or discipline, gamification features, research question(s), hypothesis/es, participant data (gender, age), independent and dependent variables and their groups/levels, measurement (subjective and objective), data source, number of items, format of data or responses, validation procedures, analysis methods, and major findings. Descriptive statistics were calculated. The extracted findings on gamification removal were separated (in the case of multiple findings per study) and then categorized as positive, negative, mixed, or neutral. Descriptive statistics were then calculated for these categories.

---

[2] For instance, items and language specific to the medical domain, such as patient populations, were excluded or modified as appropriate.

[3] https://osf.io/75cqh

## 3 RESULTS

From an initial 52 papers, a small corpus of 8 papers reporting on eight studies were identified as eligible. Summarized descriptions are presented in Table 1 and Table 2. The Moldon, Strohmaier, & Wachs [29] paper was a preprint posted to arXiv. As it had not undergone peer review, its findings were not included in that analysis; it was only used in describing the nature of studies. Also, the Featherstone [10] paper did not report on the results of gamification removal, despite the stated intention and research design, and so its findings could not be reported.

### 3.1 Nature of Gamification Removal

Table 1 presents an overview of study characteristics and the gamification elements used. All studies were conducted in field. Half (4/8) had an education context. Most (6/8) used a quasi-experimental design, with non-experimental components being lack of hypotheses, lack of randomization, lack of control groups, etc. All studies employed a longitudinal design, but timeframes varied: the shortest was 2 weeks [2] and the longest was 3 years [29]. Four studies included a baseline, which varied in length: 9.5 months [42], 5 months [15], and 3 weeks [10,33]. One [44] used a pre-test as a baseline. Gamification was in place for different lengths of time: 7 weeks [44], 6 weeks [10], 2 weeks [12,33,42], and 1 week [2]. Notably, Günther, Kacperski, & Krems [15] applied the same gamification feature at two different times, each for 3 months. The timing is unknown in Moldon et al. [29]. A clear research question was found in only 4/8 papers, but all stated the purposes for gamifying and studying removal. Reasons for gamifying included increasing user activity [2,15,29,42], improving performance [33,44], enhancing experience [10], and understanding the effect on motivation [12]. For removal, reasons included isolating the influence of gamification on behavior [2,10,29,42,44] and determining maintenance of effects over time [12,15,33]. Only 2 studies included hypotheses, despite the use of experimental approaches and inferential analyses.

Most studies did not report detailed participant demographics; only Günther et al. [15] provided gender distributions and age details (mean, standard deviation, and range). 2/8 reported only the total subject pool. Gamification elements included points (5/8), leaderboards (5/8), progress indicators (5/8), rewards (3/8), badges (2/8), avatars (2/8), currency (2/8), and missions (2/8). Gamification removal involved the full removal of all gamification elements except in the case of Featherstone [10], wherein only the battle game was removed (badges, progress bars, avatars, currency, and rewards remained).

Table 2 presents an overview of variables, measures, and measurement, as well as the valence of the results (positive, negative, mixed, or leaning). Most studies used quantitative data, i.e., logs (6/8). The exception was García Iruela et al. [12], who used a subjective self-report questionnaire. All relied on quantitative data analysis, including inferential testing. Most studies (5/8) looked at presence or absence, with 3 considering order effects, and 1 considering multiple timeframes. Most studies (6/8) focused on measures related to evaluating the worth of gamification for the platform creators or purveyors, such as participation rates (3/8), performance (1/8), and app usage (1/8). User-centered alternatives included motivation for the user and course grading.

### 3.2 Findings on Gamification Removal

Most studies reported negative [2,12,42] or mixed results [12,15,44]. For instance, Thom et al. [42] found that contributions of photos and lists decreased, as did comments across the platform, and high-scorers stopped leaving as many terse greetings. In García Iruela et al. [12], perceived confidence was higher in the first two weeks of the gamified group compared to the ungamified group, but this flattened on gamification removal. In contrast, de Paula Porto et al. [33] found that gamification increased the speed of time sheet entry, but speeds dropped by over half after removal (1.99 vs .47 vs. .8). Günther et al. [15] also found that the positive effects of gamification remained, but with a slight, non-statistically significant decline. Van Nulan et al. [44] found that exam scores and course grades remained high after gamification removal, with both early- and late-starters performing better to the non-gamified group. Given that three strategies (feedback, gamification, financial rewards) were cumulatively included over time, and the inclusion of two gamification periods and removals, it is difficult to conclude what the role of gamification was. Finally, as noted, Featherstone [10] did not report on the results of removal, so the impact is unknown.

## 4 DISCUSSION

Despite the years gone by, gamification removal is still in its infancy. The current body of work is defined by a mixture of strengths and weaknesses. Methodological limitations (such as a focus on quantitative measures) and weaknesses (such as poor reporting) abound. But a number of strengths can be seen, as well. In contrast to the greater portion of HCI and UX research [36], all studies took place in the field. This echoes a turn towards research in real contexts within gamification broadly [35]. Findings on gamification removal were mixed but lean negative. Most confirm earlier fears, with even initial positive results declining on removal. Given the small pool and noted issues, firm conclusions cannot be drawn. More work needs to be done to strengthen and expand—and potentially refute—these findings. As such, instead of a typical discussion, I offer a more useful format: a research agenda.

### 4.1 Research Agenda

*4.1.1 Encore and extend: Experimentally remove gamification.* In essence, this is a repeat of the original call [40] to examine the repercussions of gamification removal. But it is also an extension of that call, with a view to going beyond the low-hanging fruit. While some of the below may have been applied to or suggested for other gamification topics, they also need to be applied to gamification removal.

*Research designs:* Experiments—the more tightly controlled, the better—with comparative designs will be needed. Three designs are recommended. A pre-post design, where gamification is removed for a single group. A cross-over design, as in Van Nuland [44] and García Iruela et al. [12], where the gamified and ungamified groups are switched mid-study. And designs with a control group, which can be those who do not experience the gamified system or those who do not have the gamification elements removed. Importantly, these designs should pinpoint the effects of gamification elements *and* their removal. A cumulative design, such as in Günther et al. [15], may allow for a more natural and multifaceted result, but also prevents us from knowing the exact effects of gamification itself. Research questions should be clearly stated (3/8 did not) and hypotheses should be used/stated (only 2/8 did).

*Variables, measures, and measurement*: Gamification removal may be multifaceted, and so the variables under study and how they are measured should reflect that. Possible individual level and group level variables should be considered. For instance, in Thom et al. [42], country mediated responses to gamification and its removal in the same system. Other possibilities from the extant literature include personality [21], game-oriented user types [43], and orientations to motivation [16], to name a few. Selecting individuals as exemplars and extreme cases for case studies, as in Featherstone [10], may help illustrate differences. The use of logs and other noninterfering and objective methods of data collection should continue. While increasing in popularity, we should be cautious about using self-reporting methods *within* gamified systems, as gamification itself can influence reporting [14].

*Data collection methods and analysis:* Quantitative methods can identify the presence and extent of effects; the current body of work does this effectively, and this should continue. However, *why* there is or is no effect needs to be addressed more effectively. Thom et al. [42], for example, identified a difference based on country, but did not conduct research on the reason(s) for it. Speculation and comparisons to similar research are a good starting point but not sufficient to explain the specific case. Qualitative methods can be used to understand the "why" and provide context where the quantitative cannot; this is where the current body of work can be improved. Also, *when* data is collected is paramount. Measuring at the point of removal may be too soon to see effects. García Iruela et al. [12], for instance, show a likely novelty effect. Stagger when gamification is removed and when effects are measured, as in de Porta et al. [33] and Günther et al. [15]. A cross-over design is a simple but rigorous option [23]. *Report all results:* Reporting the effects of removal—not just the *act* of removal, as in Featherstone [10]—will be necessary. Negative results add to knowledge just as positive results do [28]. Further, a reader cannot assume a negative result when *no* result is reported. Negative results can resolve disputes and point to the research trajectories that are worth pursuing, as well as clarify the effects of gamification removal itself.

*4.1.2 Interrogate the negative results.* The surveyed papers, theoretical background, and other review work have highlighted the possibility of negative results upon gamification removal. We should not shy away from this [20]. Yet, the meaning of these "negative" results needs reconsideration and deeper interpretation. I outline some possibilities here.

Table 1: Study and gamification characteristics; papers sorted by year.

| Source | Area | Research Design | Timeframe | Participants | Gamification Elements |
|---|---|---|---|---|---|
| Thom, Millen, & DiMicco (2012) [42] | Workplace | Between-subjects quasi-experiment in field | 10 months of use; data collected 2 weeks before and 2 weeks after use (4 weeks total) | 3486 employees (1815 in the US, 287 in India) | Points, leaderboards; all removed |
| Amriani, Aji, Utomo, & Junus (2014) [2] | Education | Between- and within-subjects quasi-experiment in field | 2 weeks of use; 1 week gamified or not gamified; order switched between groups | 38 high school students aged 16-18: 14 men and 24 women | Points, leaderboards, badges ("titles"), progress bar ("completion track"); all removed |
| Featherstone (2017) [10] | Education | Within-subjects quasi-experiment in field | 12 weeks of use; 3 weeks off, 6 weeks on, then 3 weeks off | 26 2nd-year undergrads | Badges, progress bar, avatars, currency, rewards, battle; only battle activity removed |
| de Paula Porto, Ferrari, & Fabbri (2019) [33] | Workplace | Within-subjects quasi-experiment in field | 7 weeks of use; 3 weeks of baseline, 2 weeks on, then 2 weeks off | 13 employees in a software development company aged 21-29: 10 men and 3 women | Points, leaderboards, progress bar, avatars, currency ("gold"), rewards, missions, health bar, guilds; all removed |
| Günther, Kacperski, & Krems (2020) [15] | Energy | Within-subjects field study | 22 months of use; 5-month baseline, 3 months for each of Feedback, Gamification I, Rewards, Gamification II, then 5 months off | 108 university staff aged 18-64 (M= 34, SD=8.9); 89 men and 28 women | Leaderboards |
| García Iruela, Fonseca, Hijón-Neira, & Chambel (2020) [12] | Education | Within-subjects quasi-experiment in field | 4 weeks of use; 2 weeks gamified or not gamified; order switched between groups | 169 to 113 (77 dropouts) 1st-year undergrads; mean age of 20 | Points, leaderboards, progress bar, rewards, missions, time limits; all removed |
| Moldon, Strohmaier, & Wachs (2020) [29] | Software Development | Within-subjects field study | 3 years; ~1.5 years on, then removed | 433,138 GitHub software developers | Points (# of days streak), progress (# days in current streak); all removed |
| Van Nuland, Roach, Wilson, & Belliveau (2014) [44] | Education | Within-subjects experiment | 13 weeks; switched at week 7 | 67 2nd-year undergrads | Competitive quiz; removed |

Table 2: Variables, measurement, and valence of findings; papers sorted by year.

| Source | IV | DV | Measure | Measurement | Valence |
| --- | --- | --- | --- | --- | --- |
| Thom et al. [42] | Country (US vs. Indian) | Participation | Objective measures: # of photos, # of lists, # of comments, # of terse greetings | Logs, # of type of comments | Negative |
| Amriani et al. [2] | Presence or absence of gamification; order of the gamified week | Participation | Objective measures: Course view, assignment submission, discussion created, discussion reply and discussion view | Logs | Negative |
| Featherstone [10] | Presence or absence of battle games | App usage | Objective measures: attendance, handing in assignments, completing tutorials, manual awards from staff based on spontaneous activities (such as answering a question) | Logs | n/a[a] |
| de Paula Porto et al. [33] | Presence or absence of gamification | Performance | Objective measures: Delay in completing time sheet (# of days) | Logs | Positive |
| Günther et al. [15] | Intervention phase (cumulative) | Eco-driving behavior | Objective measures: Average energy consumption while driving | Logs | Mixed-Positive |
| García Iruela et al. [12] | Order of the gamified week | Motivation | Subjective measure: items from the Intrinsic Motivation Inventory (IMI) [6], including pressure/tension, perceived choice, perceived competence, effort/importance | Questionnaire | Mixed-Negative |
| Moldon et al. [29] | Presence or absence of gamification; streakers and non-streakers | Participation | Objective measures: # of streaks, length of streaks (days), activity on weekends (share of activity), streak after 100-day goal, streaks in social networks based on homophily (similarity) | Logs | Negative[a] |
| Van Nuland et al. [44] | Presence or absence of gamification | Performance | Objective measures: Exam scores (mid and final), final grade | Exams | Mixed-Positive |

[a] Findings may be inconclusive due to poor research design, analysis, and/or the pre-print, pre-peer review status of the paper.

*Engage with theory*: A weakness of older gamification research (cf. [31,35,40]) was a disconnect between theory and the phenomena under study. Over time this has improved for certain measures, especially motivation, and certain theories, especially Self-Determination Theory (SDT) [31]. But we need to consider other factors. For instance, some work suggests that *forcing* fun and games on people can make an enjoyable activity a negative one [7,18,30]. Yet, a reduction in performance following gamification removal may be interpreted in different ways. For instance, if the user perceived the gamification elements as extrinsic, declines may be due to a natural regression to the mean. But if they were intrinsically enjoyable, declines could be related to declines in satisfaction, even momentary ones. Even further, if long-term exposure to a gamified system leads to "offloading self-regulation" [7], taking it away could leave a person dazed and confused, without direction or the ability to self-manage.

*Specify the recipient(s) of the negative effects*: We should also distinguish for whom the result is negative: those employing gamification, hoping to harness its benefits by influencing or controlling others, or those on the receiving end, who may be extrinsically or intrinsically motivated. When it comes to extrinsic motivation, those that no longer feel the crack of the "electronic whip" [1,7] may be less inclined to perform according to the powers that be—and may even be happier. We will need to rethink what is being measured and for the benefit of whom, users *and* deployers. Self-reports on satisfaction, affect, joy, stress, and other user-centered measures could counterbalance the objective user performance measures most studies relied on, collected for the benefit of the overseer.

*Clarify the source*: Outcomes can have one or more sources (see 4.1.1) or be *confounded* by other factors. For instance, the cumulative effects and/or interactions between the incentive measures applied in Günther et al. [15] could have influenced the effects of gamification before and after removal. Comparisons of different gamification elements and/or the removal of *single* elements will be needed, as in Featherstone [10]. Studies will need to be replicated in different contexts to pinpoint source(s), such as points in an e-health app compared to an online class. Known confounding factors, such as usability, system errors, technical ability, and attitudes towards games should also be considered. These may be specific to the context, such as driving experience in Günther et al. [15].

*4.1.3 Harness the wild.* An accessible and ecologically valid way to answer this call may be found "in the wild." Half of studies took place in an educational context, likely because many of us are within or adjacent to such contexts. We should continue to explore learning environments so as to generate points of comparison. But companies and organizations are also continuing to invest in gamification [3]. Many platforms are public. As Moldon et al. [29] point out, websites like StackOverflow even change gamification features over time, leading to publicly visible effects on user behavior. Users also share their reactions on public forums, especially social media. For instance, a 2016 Nike redesign of their gamified training app prompted public outcry [45]. The removal of gamification elements was not well-received by such a large portion of the user base that Nike responded with various conciliatory measures and a fix. Using such data is robust [34,39]. Gamification researchers can use this as a data source by identifying when platforms have added and removed gamification features, even working with platform creators to do so.

*4.1.4 Applicability to gamification research in general.* This research agenda can apply to all gamification work going forward. The issue of removal is fundamental: in most research on gamification, people are subjected to gamification over some period of time, and then it is (presumably) removed. Short-term work and one-off studies may be exempt, but this needs to be studied. Future and ongoing work can incorporate modifications to research protocols, while previous studies can be re-run.

## 5 CONCLUSION

Gamification research and practice has advanced since the topic of gamification removal was first raised. As this rapid review indicates, there is a risk of reversals in positive outcomes and other negative effects, as well as positive ones. The sheer amount of research on gamification and gamification platforms in use out in the world demands our attention. At this time, there is a small corpus from which to draw conclusions, one subject to methodological weaknesses and limitations. We can move forward by acknowledging this and tackling a new research agenda. We can and must do better to understand the real effects of gamification removal in all their shades of grey.

## ACKNOWLEDGMENTS

My sincere gratitude to Peter Pennefather for his guidance and early feedback. Thank you also to the reviewers.